# Biophysical Sequence Analysis of Functional Differences of Piezo1 and Piezo2

J. C. Phillips

Dept. of Physics and Astronomy

Rutgers University, Piscataway, N. J., 08854

Abstract

Because of their large size and widespread mechanosensitive interactions the only recently discovered titled transmembrane proteins have attracted much attention. Here we present and discuss their hydropathic profiles using a new method of sequence analysis. We find large-scale similarities and differences not obtainable by conventional sequence or structural studies. These differences support the evolution-towards-criticality conjecture popular among physicists.

1. Introduction

Piezo1 and Piezo2 are ~2600 amino acid (aa) proteins that shape and reshape cell membranes between resting and functional states. They were discovered [1] as a new family of mechanically activated cation channels with strong temperature dependence [2]. They have an unusually large number (~40) of helical transmembrane regions, inferred from alternating hydropathic regions, with length ~ 21 amino acids assumed by Uniprot sequence analysis. While Piezo1 and Piezo2 have many static structural homologies when detached from membranes and studied with cryo-electron microscopy [3,4], the origin of their in vivo functional differences is unclear. A recent review noted that Piezo1 is present in nonsensory tissues, with particularly high expression in the lung, bladder, and skin; by contrast, Piezo2 is predominantly present in sensory tissues, such as dorsal root ganglia (DRG) sensory neurons [5]. The authors also state that "whether

Piezo2 and Piezo1 possess similar structures and mechano-gating mechanisms remains unknown".

A new biophysical method of sequence analysis has recently quantified the connection between sequence mutations of the Coronavirus (CoV) spike and its evolving contagiousness with high accuracy [6-9]. The method is based on a conjectured connection between living systems and evolution by natural selection towards a thermodynamic critical point [10,11]. Piezo1 and 2 are nearly the same in mammals, which suggests that the human sequences are already very close to their critical points. Proteins are the most important examples of self-organized networks, and the mechanical properties of such networks near critical points, have been extensively studied in primarily covalently bonded network glass alloys [12]. The high accuracy of the biophysical results for functional evolution of CoV spikes has resulted from the long, slender spike structures' immersion in water [6-9]. Similarly several of the functional differences between Piezo1 and 2 can be explained by profiling differences in their interactions with water, as we shall see.

2. Methods

Our earlier analysis relied on $\Psi(aa,W)$ hydropathic profiles, where $\Psi(aa)$ measures the hydropathicity of each amino acid [13,14,7]. Protein shapes are economically described by structural domains, which are known in detail for static structures. In general these shapes change slightly when attachment occurs. Here a surprising simplification occurs, which is discussed at length in earlier CoV papers [6-9]. It is a common experience that the surf of water waves near an ocean shore is larger when there is a strong wind. Similarly, small changes in protein shapes are often driven by waves in water films. It has long been thought in 20th century molecular dynamics simulations (MDS) that the interactions of water molecules with individual amino acids are complex, as indeed they are when approached from their multi-parameter Newtonian interatomic force-field perspective. An enormous simplification occurs when the 17th century Newtonian view is replace by 18th century thermodynamics and 19th century wave descriptions in the 21st century structure-based methods described here. Evolution through natural selection finds the mutations that increase attachment rates of spike viruses through water-wave driven domain synchronization [6].

These water waves have been averaged linearly over sliding windows of width W. (Data processing using sliding window algorithms is a general smoothing and sorting technique discussed online.) A natural choice for W in transmembrane (TM) proteins is 21, as used by Uniprot in listing TM segments of Piezo1 and 2. Here we use W = 17, as it appears to give higher resolution. Results were also tested using the scales appropriate to first (KD) - and second (MZ) - order transitions [13,14]. The transition from outside to inside the membrane should be first order, and tests showed more consistent results with the KD scale.

A practical aspect in obtaining $\Psi(aa,W)$ hydropathic profiles is that this matrix is displayed on a computer as a spreadsheet. EXCEL is the most common tool used for spreadsheets, and it contains a number of subroutines that are useful in identifying an optimal value for W. In preparing such an EXCEL macro, one must use their famous lookup table to assign values to each $\Psi(aa)$.

Results

The hydropathic profiles of Piezo1 and 2 are shown in Figs. 1 and 2. In both cases the "snake" is separated into two parts by a central hydrophilic valley. Cryo-electron microscopy studies of dry samples detached from membranes have shown a static trimeric structure of "blades" joined by "beams" at the C terminals [3,4]. The central hydrophilic valley is apparent structurally from its absence of TM helices in the cryo-electron microscopy structures [3,4]. The "beams" correspond roughly to its N-terminal half. The TM segments in the figures are numbered as listed by Uniprot Q92508 and Q9H515 from their sequence analysis. The profiles are similar, but they also contain interesting differences.

The wide hydrophilic valley is stabilized by a single TM segment in Piezo2 centered at 1640, and by three closely spaced TM segments in Piezo1 (numbered 26-28 in Fig. 1, centered at 1695, 1712 and 1741). The Piezo1 hydrophobic maxima lie in the range 190-215, and the three central maxima are near 200. The Piezo2 hydrophobic maxima lie in the range 195-220, with the single central maximum, labeled !! in Fig. 2, at 193.5. Uniprot does not list a TM segment here, but it does list three TM segments near 2320 with values near 190. These differences in sequence analysis appear to be minor, except that the single central maximum in Piezo1 is important in the analysis made below.

In addition to the nearly level sets of hydrophobic maxima associated with TM segments, there are hydrophilic extrema that span a wide range, from hydroneutral (near 155) to very strongly hydrophilic (near 100). Some of these are marked in the Figures. In Piezo1 the beam (C terminal) region has one strongly hydrophilic extrema, while the blade region has two. They divide the TM segments into groups. In Piezo2 the blade region has added a strongly hydrophilic extremum in the blade region. The strongly hydrophilic extrema all lie in a narrow range near 100.

We can now see how these differences alter the dynamical properties of Piezo2 compared to Piezo1. The central hydrophilic valley acts as a hinge between the beam and the blade. Its three TM segments have shrunk to one TM segment in Piezo2, making Piezo2 more flexible. Also the addition of an extra strongly hydrophilic extremum to the Piezo2 blade makes it more flexible. Thus overall Piezo1 is more stable, while Piezo2 is more flexible.

Discussion

In comparatively simple systems (like covalent glass alloys), it was possible to measure the elastic properties of the glass networks directly and plot them against force field constraints [12]. In proteins this is seldom the case, even for proteins smaller than 300 amino acids. At the opposite extreme of the ~ 2600 aa Piezo proteins this is not possible. However, given their function as mechanotransducers, one expects that their two (and only two) types could have slightly different elastic properties, and that both are close to an evolutionary critical point.

Signs of the antiquity of Piezo1 (more stable) are apparent from peculiar aa repetitions. Near 745 one sees EEQQEHQQQQQEEEEEEE. This unique 18 aa sequence of glutamic acid (E) and glutamine (Q) is responsible for the Smin1 hydrophilic extremum in Fig.1. Both MZ and KD scale values [7] are nearly equal for these two aa. For the hydrophobic extrema we can look at the first 40 aa of Prot1, which contains two closely spaced TM segments. Here we find 16 Leu (40%!), and the success of the KD scale in finding TM segments in hydrophobic extrema can be traced to its larger value of $\Psi$(Leu). (Overall Leu is 15% of Piezo1, and only 11% of Piezo2.) However, we prefer to regard the KD scale as being directly connected to the thermodynamically first order water-air transition. The advantage of scaling [10,11] is apparent here, as profiles goes far beyond amino acid counting.

Evolution refined Piezo1 into Piezo2 in Fig.2. There is a refined Smin1c near 463, with the 21 aa sequence EKREEEEEEKEEFEEERSREE, including 15 glutamic acids E and no glutamine Q (replaced by more hydrophilic K and R). Similarly EESEEDGEEEEESEEEE is associated with Smin1a near 890.

Further evidence showing that Piezo2 is more flexible than Piezo1 is obtained by counting Proline fractions. Proline is the only amino acid with a double connection to the peptide backbone, which is why Pro pairs have been so useful in stabilizing Spike vaccines [17]. The fraction of Pro in Piezo2 is 0.040, which increases in Piezo1 to 0.056 (40% increase). The double connection of Pro is the simplest example of constraint theory [12], and its success here shows that the dominant bonding is covalent.

One can go further by studying Pro pairs. In Piezo1 a random distribution would produce 7.8 Pro pairs, while there are only 3 pairs, at 276, 1376, and 2282. In Piezo2 random would give only 4.4 pairs, while there are 8 pairs, at 300, at 1188, and then six at 1588, 1830, 1873, 2003, 2511, and 2660. Thus Piezo2 is more refined, has formed Pro pairs, and is using them to stabilize long-range interactions, especially with six in the half nearer the C terminal.

Differences between Piezo1 and Piezo2 mechanofunctions (summarized in [5]) were dramatically identified by studying chondrocytes, the cells in cartilage [15]. The cartilage is taken from diarthrodial joints, where they sustain millions of cycles of mechanical loading. As they show in their Fig.1, the levels of Piezo1 and Piezo2 are similar in lungs, bladder and skin. Piezo1 levels are high in cartilage, and low in evolved neuronal tissue, while the reverse holds for Piezo2. Further studies showed that PIEZO2 was highly expressed in DRG neurons of all sizes, while PIEZO1 was selectively expressed in smaller neurons [16].

In general one would expect more stable Piezo1 to appear in cartilage [15]. Because Piezo2 is more flexible, it would be able to reshape neuronal cells more rapidly in the context of neuronal network signals. Moreover, as Piezo1 is more stable, its effectiveness would be limited to smaller neurons [16]. The existence of two very large homologous and more stable transmembrane proteins is consistent with conjectures of evolution to critical points [10,11].

While it has not been tested on sequence data from a series of candidates in a development program, one can still hope that sequence analysis could have important applications in drug development, as it has already explained the evolution of CoV contagiousness [6-9,17].

Data sources: Amino acid sequences from Uniprot Q92508 and Q9H515. Values of Ψ(aa) for the KD and MZ scales from Table 1 of [7].

Figure Captions

Fig. 1. Hydropathic profile for Piezo1. The numbered hydrophobic extrema match the transmembrane segments listed in Uniprot. Because of the protein length, the original figures were displayed on a different screen (EXCEL) twice as wide.

Fig. 2. Hydropathic profile for Piezo2. Again there is a good correspondence with Uniprot's TM segments, with the exception of the hydrophobic extremem in the center of the Central Hydrophilic Valley.

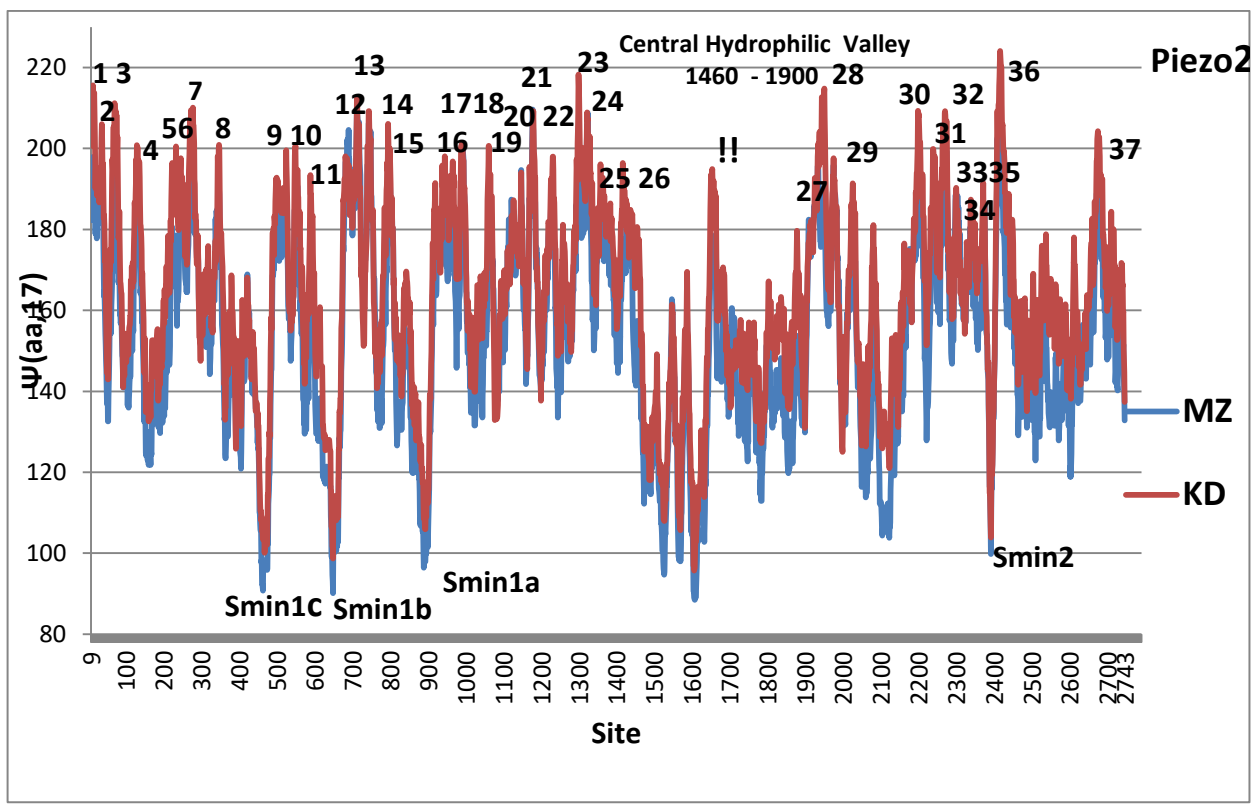

**Fig. 1.**

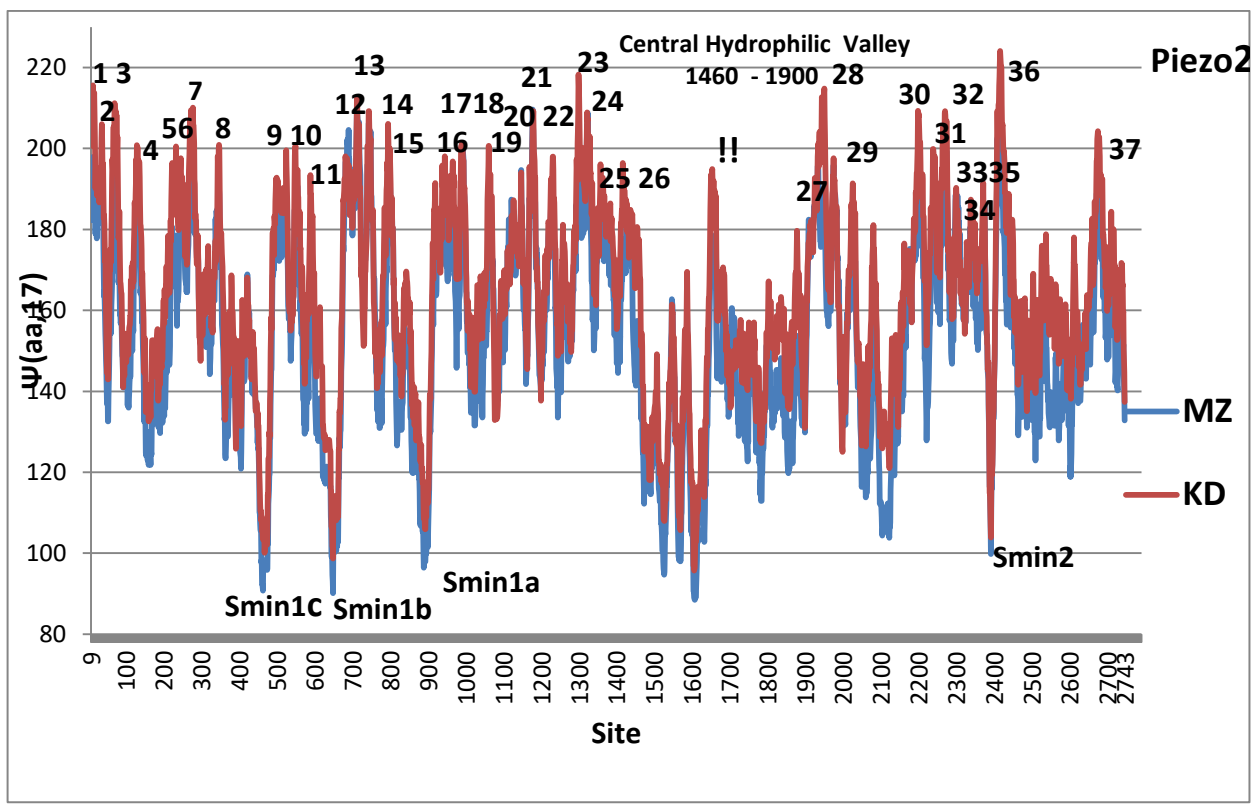

**Fig. 2.**